\documentclass[aps,prb,amsmath,amssymb,floatfix,superscriptaddress]{revtex4}

\usepackage{color}
\usepackage{tabu}
\usepackage{graphicx}
\usepackage{dcolumn}
\usepackage{bm}
\usepackage{array}
\usepackage{float}
\usepackage{supertabular}
\usepackage{longtable}
\usepackage{mathrsfs}
\usepackage{txfonts}
\usepackage{wasysym}
\usepackage{hyperref}

\begin{document}

\title{Optimal noise-canceling shortcuts to adiabaticity: application to noisy Majorana-based gates}

\author{Kyle Ritland}
\affiliation{Department of Physics and Astronomy, Western Washington University, Bellingham, Washington 98225, USA}
\author{Armin Rahmani}
\affiliation{Department of Physics and Astronomy, Western Washington University, Bellingham, Washington 98225, USA}
\affiliation{Advanced Materials Science and Engineering Center (AMSEC),Western Washington University, Bellingham, Washington 98225, USA}

\date{\today}


\begin{abstract}
Adiabatic braiding of Majorana zero modes can be used for topologically protected quantum information processing. While extremely robust to many environmental perturbations, these systems are vulnerable to noise with high-frequency components. Ironically, slower processes needed for adiabaticity allow more noise-induced excitations to accumulate, resulting in an antiadiabatic behavior that limits the precision of Majorana gates if some noise is present. In a recent publication [Phys. Rev. B 96, 075158 (2017)], fast optimal protocols were proposed as a shortcut for implementing the same unitary operation as the adiabatic braiding. These shortcuts sacrifice topological protection in the absence of noise but provide performance gains and remarkable robustness to noise due to the shorter evolution time.
Nevertheless, gates optimized for vanishing noise are suboptimal in the presence of noise. If we know the noise strength beforehand, can we design protocols optimized for the existing unavoidable noise, which will effectively correct the noise-induced errors? We address this question in the present paper.  We find such optimal protocols using simulated-annealing Monte Carlo simulations. The numerically found pulse shapes, which we fully characterize, are in agreement with Pontryagin's minimum principle, which allows us to arbitrarily improve the approximate numerical results (due to discretization and imperfect convergence) and obtain numerically exact optimal protocols. The protocols are \textit{bang-bang} (sequence of sudden quenches) for vanishing noise, but contain continuous segments in the presence of multiplicative noise due to the nonlinearity of the master equation governing the evolution. We find that such noise-optimized protocols completely eliminate the above-mentioned antiadiabatic behavior. The final error corresponding to these optimal protocols monotonically decreases with the total time (in three different regimes) and extrapolates to zero in the limit of infinite time. Our results set the precision limit of the device as a function of the noise strength and total time.

\end{abstract}
 
\maketitle
\section{Introduction}

Majorana zero modes (MZMs) are promising candidates for topological quantum computing~[\onlinecite{kitaev_fault-tolerant_2003}]. Remarkable progress has been made in the experimental realization of MZMs [\onlinecite{oreg_helical_2010,lutchyn_majorana_2010,
alicea_new_2012,beenakker_2012,elliott_majorana_2015,mourik_signatures_2012,das_zero-bias_2012,churchill_superconductor_2013,rokhinson_fractional_2012,deng_anomalous_2012,finck_anomalous_2013,nadj-perge_observation_2014,aasen-milestones-2016}]. The nonlocal nature of quantum information processing by \textit{adiabatic} braiding of MZMs makes them resilient to environmental perturbations~[\onlinecite{nayak_non-abelian_2008}]. Adiabaticity is necessary for topological protection. However, these systems are not protected from all possible errors, such as noise with high-frequency components [\onlinecite{goldstein_decay_2011,budich_failure_2012,schmidt_decoherence_2012}]. Furthermore, other sources of error, like quasiparticle poisoning~[\onlinecite{rainis_majorana_2012}] (although abated in recent designs by the charging energy~[\onlinecite{karzig2017}]), may limit the available timescales of the process, leading to unavoidable nonadiabatic effects~[\onlinecite{cheng_nonadiabatic_2011,karzig_boosting_2013,amorim_majorana_2014,karzig_shortcut_2015,zhang_shortcut_2015,karzig_optimal_2015,knap_quick_braid_2016,sekania2017}].  Last but not least, slow adiabatic dynamics is undesirable from the performance perspective.

In recent years, significant progress has been made in the emerging area of \textit{shortcuts to adiabaticity}~[\onlinecite{chen_fast_2010,torrontegui2013,jarzynski2013}], namely, performing dynamical quantum operations which yield the same outcome as the adiabatic evolution but in a shorter time. One of the approaches to finding such shortcuts is through \textit{optimal control} [\onlinecite{peirce_optimal_1988,palao_quantum_2002,kral_coherently_2007,hohenester_optimal_2007,salamon_maximum_2009,rahmani_optimal_2011,doria_optimal_2011,caneva_speeding_2011,choi_optimized_2011,del_campo_fast_2011,hoffmann_time-optimal_2011,rahmani_cooling_2013}]. An optimally fast scheme was derived in the absence of noise in Ref. [\onlinecite{rahmani_optimal_2017}], which produces the same unitary operation as adiabatic (infinite-time) braiding of MZMs in a short time. While sacrificing strict topological protection, this scheme offers practical advantages in balancing performance and robustness. The advantages are especially important when the control fields are noisy.

Noise, which may contain high-frequency components, is generically unavoidable and not protected against by topology. Furthermore, noise-induced errors accumulate over time and become more harmful in the adiabatic limit. Hence, the fast protocols of Ref. [\onlinecite{rahmani_optimal_2017}] suffer from much fewer noise-induced errors than slow adiabatic protocols that take a long time. In fact, noise results in an antiadiabatic behavior, where slowing down the dynamics can actually increase the diabatic excitations~[\onlinecite{dutta_anti_2016}].

Nevertheless, the protocols of Ref. [\onlinecite{rahmani_optimal_2017}] are designed assuming vanishing noise, and suffer from a similar noise-induced heating \textit{rate} characteristic of adiabatic protocols. This heating is limited since the fast protocols expose the system to noise for a shorter time than adiabatic protocols. In this paper, we obtain protocols that are optimal for an a priori known and unavoidable noise. They are optimized to \textit{cancel out the effects of the existing noise}. Finding these noise-canceling optimal shortcuts helps us determine the \textit{precision limit} of a Majorana-based gate in the presence of noise, as the protocols represent the best we can do given a certain total time, an existing noise strength, and an experimentally relevant range for the tunable inter-Majorana hybridization energies.

Throughout the paper, we use multiplicative noise, which is dominant for topological qubits. We focus on white noise for simplicity. A ubiquitous source of noise in superconducting devices, namely, the pink $1/f$ noise, has a slowly decaying noise spectrum. It was shown in Ref. [\onlinecite{rahmani_optimal_2017}] that pink and white noise have a qualitatively similar antiadiabatic effect on topological gates.

We address several open questions: What is the minimum error we can get if the protocols are optimized for the existing strength of noise? Is exact preparation of the unitary corresponding to the adiabatic evolution possible in the presence of noise? Will the antiadiabatic effect still occur when using noise-canceling optimal protocols? To answer these questions, we focus on a widely used minimal effective model of MZM braiding~[\onlinecite{hassler_toptransmon_2011,
vanheck_coulomb_2012, hyart_flux_2013}], whose evolution is governed by a Lindblad master equation [\onlinecite{pilcher_heating_2013,rahmani_quantum_2013,rahmani_dynamics_2015,dutta_anti_2016}]. We find optimal protocols by numerical minimization (through simulated-annealing Monte Carlo simulations) of a cost function based on the trace distance of the noise-averaged final density matrix from the perfectly adiabatic target density matrix. 

We also perform an in-depth analysis of the mathematical optimal-control theory of our system. In agreement with Pontryagin's minimum principle~[\onlinecite{pontryagin_mathematical_1987}], we find that our protocols are bang-bang for vanishing noise (and close to bang-bang for weak noise). Upon increasing the noise strength, protocols with non-bang-bang segments occur due to the nonlinearities of multiplicative noise, still in full agreement with Pontryagin's principle. We fully characterize these optimal protocols. The connection to Pontryagin's principle is an important check that the Monte Carlo simulations have indeed found the optimal protocol. It can also help eliminate the discretization errors and imperfect convergence in the simulations, yielding numerically exact optimal protocols.
Furthermore, in the regime of weak noise and short times, where Monte Carlo has divergence issues, a much more efficient bang-bang ansatz is used, which can yield very high-performance protocols. 

Our main results are as follows. As a function of total time, we find three regimes in the behavior of the minimum cost function (obtained from optimal protocols customized for each total time and noise strength). In the first regime, the cost function does not change from its initial value, implying there is \textit{no} permissible evolution with a total time smaller than a critical time $\tau_c$ that can reduce the cost function. In the second regime, the cost function rapidly decreases as we increase the total time (in the absence of noise, this rapid decent continues all the way to zero), and in the third regime, the cost function exhibits a slow decay for finite noise (for vanishing noise we have a plateau with a vanishing cost function). Extrapolating the cost function of the third regime to the infinite-time limit, we find the cost function monotonically decreases with the total time $\tau$ and eventually vanishes as $\tau\to \infty$. Antiadiabatic behavior is completely eliminated by using these noise-canceling optimal protocols. Other significant contributions of the paper are the full characterization of the optimal pulse shapes. The patterns that emerge in the finite-time studies can inform numerical simulations for longer total times, where the computations become more expensive.

The outline of the paper is  as follows. In Sec. II, we describe the effective model of the system and the unitary operator corresponding to perfect adiabatic braiding. We also present an error function, suitable for noisy systems, for the deviations from adiabaticity.
In Sec. III, we discuss the Lindblad master equation used for the evolution of the noise-averaged density matrix in the presence of multiplicative (relevant to Majorana-based gates) white noise. We also discuss the antiadiabatic behavior of the system when standard protocols are used [see Fig.~\ref{fig:schem}(b)]. In Sec. IV, we discuss the Monte Carlo and bang-bang optimization methods, and present our numerical results for the minimum error function as a function of noise strength and total time, as well as its extrapolation to the infinite-time limit. We find three distinct regimes in the behavior of the cost function. Sec. V is focused on the application of the mathematical theory of optimal control and Pontryagin's minimum principle to our system. In terms of the time-dependent density matrix and certain conjugate momenta, we derive the analytical form of the optimal protocols. In Sec. VI, we verify our numerically found optimal protocols against the analytical results of Sec. V, and characterize the patterns of the optimal protocols in different regimes. We close the paper in Sec. VII with brief conclusions. Some details of the application of Pontryagin's minimum principle are presented in two appendices.


\section{Model of a Majorana-based gate and its error function}
\label{sec:effective}
\begin{figure}[]
\includegraphics[width=10cm]{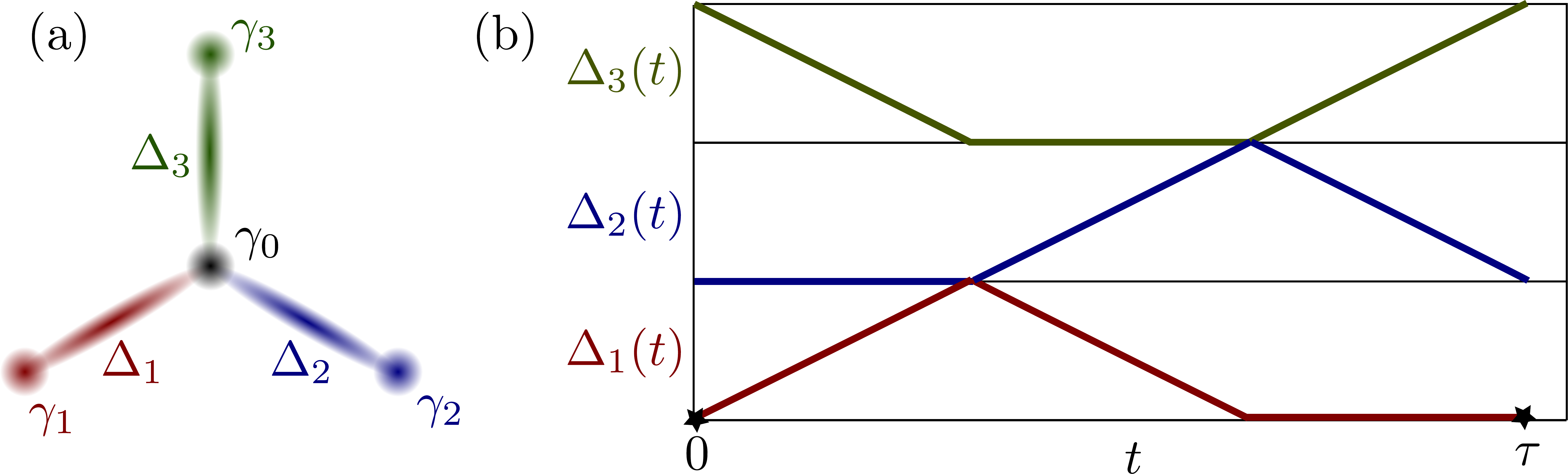}
\caption{(a) A schematic representation of the Hamiltonian~\eqref{eq:hamil0} with three Majorana modes hybridized with a central Majorana mode $\gamma_0$. The hybridization energies $\Delta_j$ are tunable. (b) Adiabatic exchange of initially decoupled $\gamma_1$ and $\gamma_2$, i.e., $\Delta_1(0)= \Delta_2(0)=0$, can be achieved by, e.g., the linear protocol above in the limit of $\tau\to\infty$.}
	\label{fig:schem}
\end{figure}
We use an effective model of Majorana braiding relevant to devices such as the top-transmon~ [\onlinecite{hassler_toptransmon_2011,
vanheck_coulomb_2012, hyart_flux_2013}]. The Hamiltonian can be written in terms of four Majorana operators as
\begin{equation}\label{eq:hamil0}
H(t)=i\gamma_0\sum_{j=1}^3\Delta_j(t)\gamma_j,
\end{equation}
where $\gamma_j=\gamma^\dagger_j$ and $\{\gamma_i,\gamma_j\}=2\delta_{ij}$. Three Majorana operators can hybridize with one central operator $\gamma_0$ as in Fig.~\ref{fig:schem}(a). Each $\Delta_j$ represents the hybridization energy between $\gamma_0$ and $\gamma_j$ for $j=1,2,3$. We assume we have time-dependent control over these three Hamiltonian parameters and can tune each of them to an arbitrary  value in the range 
$
0\leq\Delta_j(t)\leq{\cal D}$.
Hereafter, we set the maximum hybridization energy scale ${\cal D}=1$.

By defining two Dirac fermions $c=(\gamma_1+i\gamma_2)/2$ and $d=(\gamma_0+i\gamma_3)/2$,
we can write the Hamiltonian as a $4\times 4$ matrix in the basis $\left(|0\rangle, d^\dagger |1\rangle, |1\rangle, d^\dagger |0\rangle\right)$, where $ |1\rangle\equiv c^\dagger |0\rangle$ and $|0\rangle$ is the vacuum. The first (last) two basis states have even (odd) fermion parity, which is a symmetry of the Hamiltonian. We then represent $H(t)$ by the block-diagonal matrix
\begin{equation}\label{eq:hamil1}
{ H}=\Delta_1O_1+\Delta_2O_2+\Delta_3O_3,
\end{equation}
with 
\begin{equation}\label{eq:O}
O_1=\openone\otimes\sigma_{y}, \quad O_2=-\sigma_z\otimes\sigma_x, \quad O_3-\openone\otimes\sigma_z,
\end{equation}
where $\sigma_{x,y,z}$ are the Pauli matrices and $\openone$ is the $2\times 2$ identity matrix. The  top (bottom) block corresponds to even (odd) fermion parity.

A particular gate corresponding to the exchange of Majorana particles $\gamma_1$ and $\gamma_2$  (both decoupled at the beginning and the end of the process with $\Delta_1=\Delta_2=0$) generates a unitary operation
\begin{equation}\label{eq:unitaray}
U={\rm exp}\left({\pi\over 4}\gamma_2\gamma_1\right),
\end{equation}
up to an overall phase factor, in the doubly degenerate ground-state manifold of the initial (and final) Hamiltonian. This manifold is spanned by states $|0\rangle$ and $|1\rangle$ with $U|0\rangle =|0\rangle $ and $U|1\rangle =i|1\rangle$. The standard adiabatic scheme can be represented as the following cyclic process for the vector $(\Delta_1, \Delta_2, \Delta_3)$:
 \begin{equation}
(0,0,1)\to(1,0, 0)\to(0, 1,0)\to (0,0,1).
\end{equation}
As illustrated in Fig.~\ref{fig:schem}(b), for the case of turning the coupling constants on and off as linear functions of time, each step (e.g., turning $\Delta_3$ off and $\Delta_1$ on in the first step) is performed adiabatically in the standard scheme (e.g., by increasing $\tau$). To generate the same gate in an optimal manner, we minimize an appropriate cost function (which vanishes for the above adiabatic process in the limit $\tau\to\infty$) for an arbitrary cyclic process of fixed total time $\tau$ with
 \begin{equation}
(\Delta_1(0), \Delta_2(0), \Delta_3(0))=(\Delta_1(\tau), \Delta_2(\tau), \Delta_3(\tau))=(0,0,1).
\end{equation}
All three hybridization parameters can be arbitrary functions of time with only the following constraint on the range:
 \begin{equation}
 0\leqslant \Delta_j(t)\leqslant 1, \quad 0<t<\tau, \quad j=1,2,3.
\end{equation}
The ideal gate transforms an initial superposition $c_0|0\rangle+c_1|1\rangle$ into $e^{i\phi} \left(c_0|0\rangle +ic_1|1\rangle\right)$ at time $t=\tau$, where $\phi$ is an arbitrary overall phase.

As we are interested in a noisy system, a cost function defined in terms of noise-averaged quantities would be most helpful. Furthermore, we would like our cost function to be independent of $\phi$. Density matrices are insensitive to the overall phase of the wavefunction and convenient for averaging over noise, making them a useful tool for characterizing gate precision. As in Ref. [\onlinecite{rahmani_optimal_2017}], we pick a particular equal-weight superposition $|\psi_0\rangle = \frac{1}{\sqrt{2}}(|0\rangle+|1\rangle)$ as the initial state. The initial density matrix is then given by 
 \begin{equation}\label{eq:r0}
\rho_0 = |\psi_0\rangle\langle\psi_0| = \frac{1}{2}(|0\rangle\langle0| + |0\rangle\langle1| + |1\rangle\langle0| + |1\rangle\langle1|).
\end{equation}
The target state has a $\phi$-independent density matrix $\sigma= \frac{1}{2}(|0\rangle\langle0| - i|0\rangle\langle1| + i|1\rangle\langle0| + |1\rangle\langle1|)$. We choose the trace distance between $\sigma$ and the \textit{noise-averaged} density matrix $\rho(\tau)$ as the error function of the gate:
\begin{equation}\label{eq:cost} 
C[\sigma,\rho(\tau)]\equiv \frac{1}{2}{\rm tr}\sqrt{\left[\sigma-{\rho}(\tau)\right]^2}.
\end{equation} 
Notice that, due to noise averaging, $\rho(\tau)$ generically describes a mixed state. Even though we have a pure state with $\rho(0)=\rho_0$ at time $t=0$ and the system evolves coherently for \textit{each} realization of noise.

A comment is in order regarding the choice of $\rho_0$. The Hamiltonian is block-diagonal for one realization of noise so $|0\rangle$ and $|1\rangle$ are not coupled for a single realization. If the operation transforms $\rho_0$ to $\sigma$, we do not have leakage out of the ground-state manifold and the correct relative phase is imparted to these ground states. In the limit of small $C[\sigma,\rho(\tau)]$, we expect our protocols to correctly transform arbitrary superpositions of $|0\rangle$ and $|1\rangle$, not just $|\psi_0\rangle$ defined above Eq.~\eqref{eq:r0}. It is important, however, that $\rho_0$ describes a superposition to capture the relative phase.

\section{Lindblad master equation for multiplicative white noise}
\label{sec:sys}

Our tunable control parameters are $\Delta_j(t)$. However, the physical system may experience different parameters, such as $\Delta_j^S(t)$ instead of $\Delta_j(t)$, which we try to impart to the system:
\begin{equation}
\Delta_j^S(t)=\Delta_j(t)\left[1+\epsilon_j(t)\right].
\end{equation}
We have used a \textit{multiplicative} random fluctuation to account for the topological nature of the qubits. When $\Delta_j(t)$ is suppressed due to the separation distance between two Majorana modes, the wavefunction overlap between these Majorana modes is exponentially suppressed and the noise in the hybridization parameter should be similarly suppressed. Therefore, we do not expect any additive noise for a Majorana-based topological qubit. For simplicity, we focus on white noise with 
\begin{equation}\label{eq:moment}
\overline{\epsilon_{j}(t)\epsilon_{j'}(t')}=W^2\delta_{jj'}\delta(t-t'),
\end{equation}
where $W$ is the strength of noise.  Cross-talk between different control parameters is also neglected, which is a reasonable assumption for a topological qubit. While $1/f$ noise is more realistic than white noise, it has a slow power-law decay and considerable support over high frequencies. Previous studies suggest qualitative similarities between the two noise spectra~[\onlinecite{rahmani_optimal_2017}]. White noise allows us to determine the noise-averaged density matrix by efficiently solving a single deterministic master equation and is therefore very convenient for performing optimization over the shape of the protocol, which requires solving the dynamics a large number of times. 

The master equation for the multiplicative white noise above has the following Lindblad form~[\onlinecite{rahmani_optimal_2017}]:
\begin{equation}\label{eq:master} 
\partial_t{{\rho}}(t)=i[{\rho}(t),H(t)] 
-\frac{W^2}{2}\sum_{j}\Delta_j^{2}(t)\left[\left[{\rho}(t),i\gamma_0\gamma_j\right],i\gamma_0\gamma_j\right],
\end{equation}
where $H(t)$ is given in Eq.~\eqref{eq:hamil0}. The matrix representation of the master equation in our computational basis [see Eq. \eqref{eq:hamil1}] is given by
\begin{equation}\label{eq:master2} 
\partial_t\rho=\sum_j\left\{i\Delta_j\left[\rho,O_j\right] - \frac{W^2}{2}\Delta_{j}^2\left[\left [\rho,O_j\right],O_j\right]\right\},
\end{equation}
with $O_i$ matrices defined in Eq. \eqref{eq:O}. Solving the above differential equation with the initial condition $\rho(0)=\rho_0$ [see Eq. \eqref{eq:r0}] gives the final noise-averaged density matrix at time $\tau$.
\begin{figure}[]
\includegraphics[width=9cm]{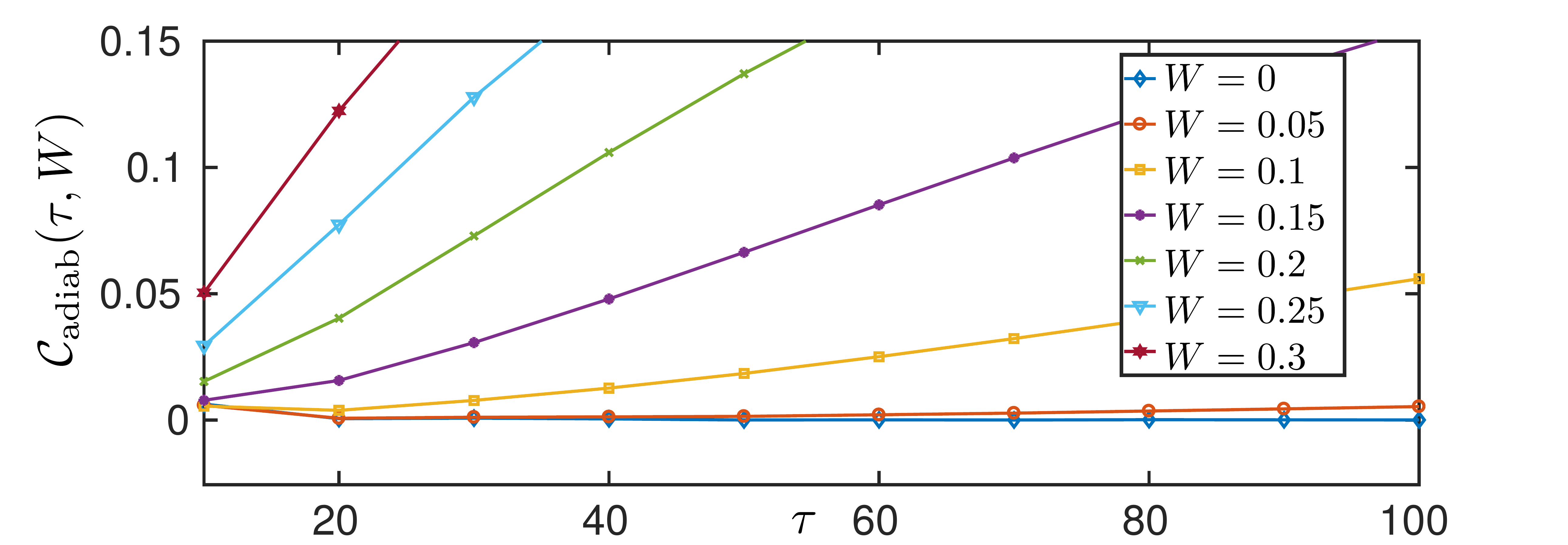}
\caption{The error function for the adiabatic exchange shown in Fig. \ref{fig:schem} as a function of $\tau$ and $W$, which exhibits an antiadiabatic behavior in the presence of noise, where slower processes (larger $\tau$) generate more nonadiabatic excitations.
	\label{fig:adiab}}
\end{figure}

Before proceeding, we calculate the above cost function [see Eq. \eqref{eq:cost}] using the linear protocol of Fig.~\ref{fig:schem}(b) in the adiabatic limit of long times $10<\tau<100$. The results are presented in Fig.~\ref{fig:adiab}. In the absence of noise, the cost function approaches zero in the limit $\tau\to\infty$. Topological protection does not save us from the detrimental effects of noise, which, ironically, accumulate over the long times required for adiabaticity. As seen in Fig.~\ref{fig:adiab}, this gives rise to an antiadiabatic behavior, where longer times lead to greater errors. The antiadiabatic behavior provides an important motivation for this work.

\section{Numerical determination of optimal protocols}
\subsection{Parameterization for simulated annealing} 
\begin{figure}[]
\includegraphics[width=6.5cm]{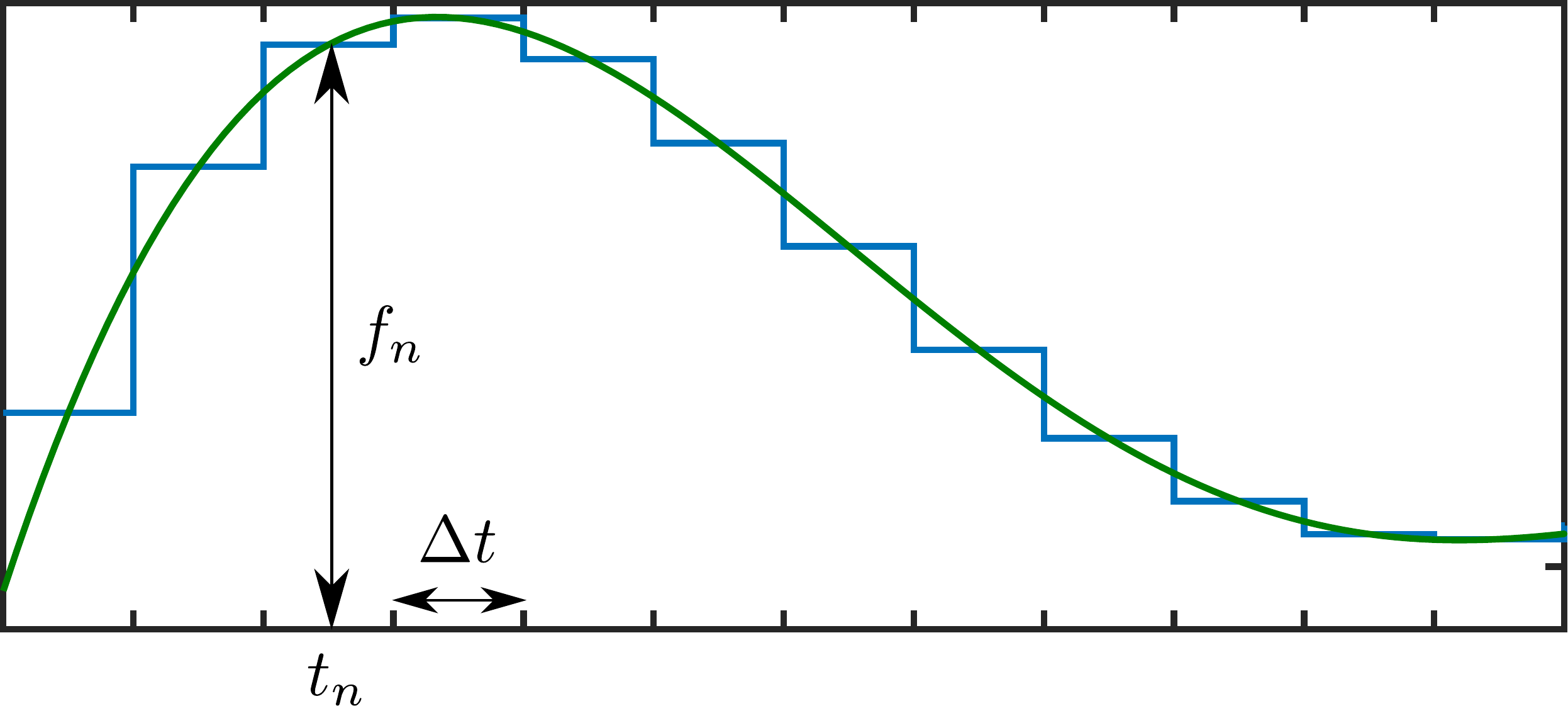}
\caption{The parameterization of arbitrary protocols by approximate piecewise-constant functions, using a discretization of time. The values $f_n$ serve as variational parameters in the Monte Carlo simulations.}
	\label{fig:pwc}
\end{figure}

Finding protocols that minimize the cost function requires a parameterization of the protocol in terms of a finite number of variables. The cost function $\cal C$ can be uniquely determined by three bounded functions $\Delta_j(t)$ for $j=1,2,3$ and $0<t<\tau$. In other words, $\cal C$ is a functional of these $\Delta_j(t)$ functions. By discretizing time, as shown in Fig.~\ref{fig:pwc}, we can approximate an arbitrary protocol as a piecewise-constant function, treating the height of these pieces ($f_n$ in Fig.~\ref{fig:pwc}) as the optimization parameters. Increasing the number of intervals allows us to scan the entire function space. Such direct minimization of the cost function is in the spirit of a machine-learning approach~[\onlinecite{bukov2017}].

We then perform simulated-annealing Monte Carlo simulations where, in each step, a random time interval $t_n$ is chosen and the three optimization parameters corresponding to the interval are varied by a small amount. If all the new values are within the permissible bounds, the move is accepted according to Metropolis rules; It it accepted with probability one if it reduces $\cal C$, (i.e., $\Delta {\cal C}<0$) and with probability $e^{-\beta \Delta {\cal C}}$ if it increases $\cal C$ ($\Delta {\cal C}>0$). The fictitious inverse temperature $\beta$ is slowly increased as the simulations run.

\subsection{Solving the master equation} 
\label{sec:master}
The simulations rely on calculating $\cal C$ for arbitrary piecewise-constant protocols in each step. By arranging the elements of the $4\times 4$ density matrix into a vector $\bar \rho$ of length 16, we can write the master equation~\eqref{eq:master} as
\begin{equation}\label{eq:master3} 
\partial_t{\bar{\rho}}(t)=K\left[W^2,\Delta_1(t),\Delta_2(t),\Delta_3(t)\right]\bar{{\rho}}(t),
\end{equation}
where $K$ is a $16\times 16$ matrix. The formal solution for constant $\Delta_j$, corresponding to each piece of the piecewise-constant protocol, is given by
\begin{equation}\label{eq:master4} 
{\bar{{\rho}}}(t+\Delta{t})=e^{K\Delta{t}}{\bar{\rho}}(t),
\end{equation}
where we have divided the total time $\tau$ into $N$ intervals of length $\Delta t=\tau/N$. The above expression leads to
\begin{equation}\label{eq:master5} 
{\bar{{\rho}}}(\tau)=e^{K_N\Delta{t}}\dots e^{K_2\Delta{t}}e^{K_1\Delta{t}}\bar{\rho}_0.
\end{equation}

%
%

\subsection{The optimal cost function from simulated annealing} 
\label{sec:brute}

\begin{figure}[]
\includegraphics[width=10cm]{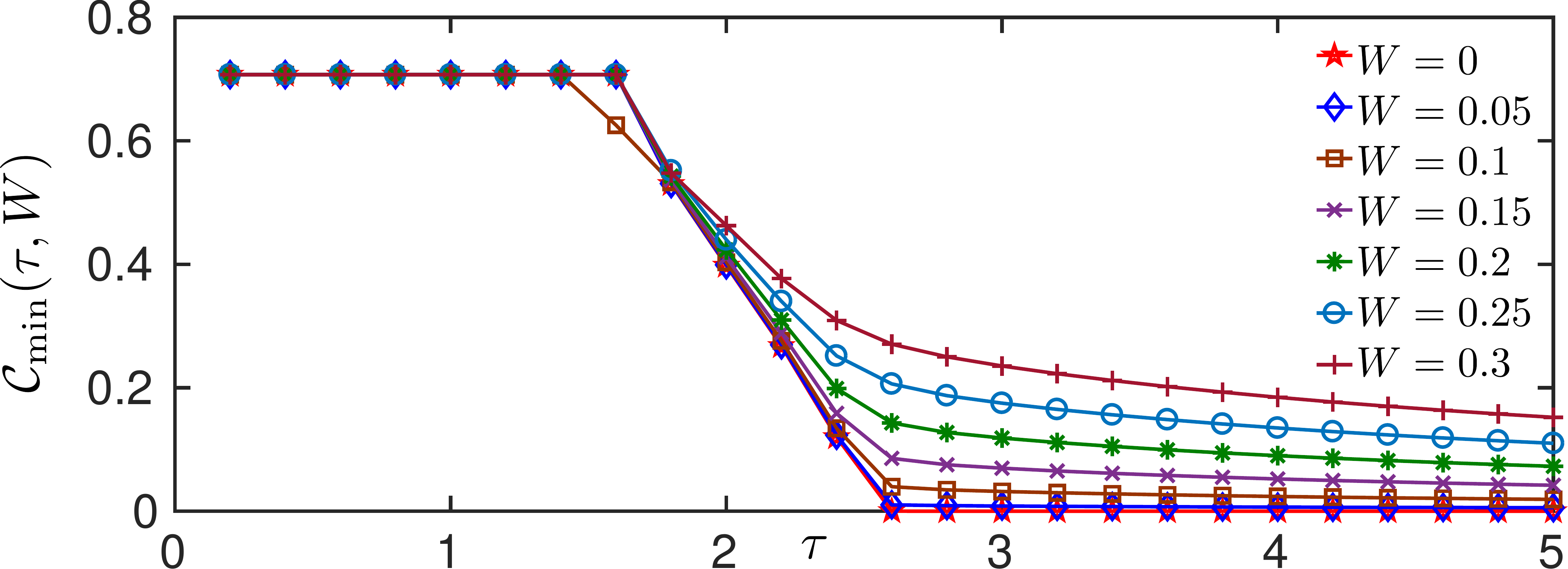}
\caption{The minimum error function for the optimal protocols obtained form Monte Carlo simulations as a function of total time. Three distinct regimes are observed. I: $\cal C$ does not decrease from its initial value. II: $\cal C$ rapidly decreases. III: $\cal C$ decays very slowly, extrapolating to zero only in the limit $\tau\to \infty$.}
	\label{fig:SA}
\end{figure}

We start our simulations by fixing $\Delta t=0.02$. The number of variational parameters is then proportional to the total time $\tau$. This direct optimization method does not make any assumptions about the shape of the protocols a priori. As it turns out, the optimal protocols for zero noise are bang-bang while the protocols for the noisy cases contain continuously changing segments. The results for the minimum cost function obtained from Monte Carlo simulations (with protocols individually optimized for each $\tau$ and $W$) are shown in Fig.~\ref{fig:SA}. We find three distinct regimes:
\begin{itemize}
\item [I:] The minimum cost function for all $W$ initially remains constant upon increasing the total time. While $\tau$ is less than a critical time $\tau_c\sim 1.3$, we cannot change the minimum cost from its initial value using quantum evolution. 
\item [II:] The second regime is characterized by the rapid decrease of the cost function as a roughly linear function of total time. For vanishing noise ($W=0$), the cost function reaches zero at the end of the second regime at $\tau\sim 2.6$. For larger strengths of noise, ${\cal C}_{\rm min}$ is finite at the end of the second regime and increases with noise.
\item [III:] Finally, a third regime occurs for longer times. For zero noise, the minimum cost function remains identically zero in this regime. For finite $W$, we observe a slow decay in $\cal C_{\rm min}$ as total time increases. The antiadiabatic limit is completely eliminated by these optimal protocols. 
\end{itemize}

\begin{figure}[]
\includegraphics[width=8cm]{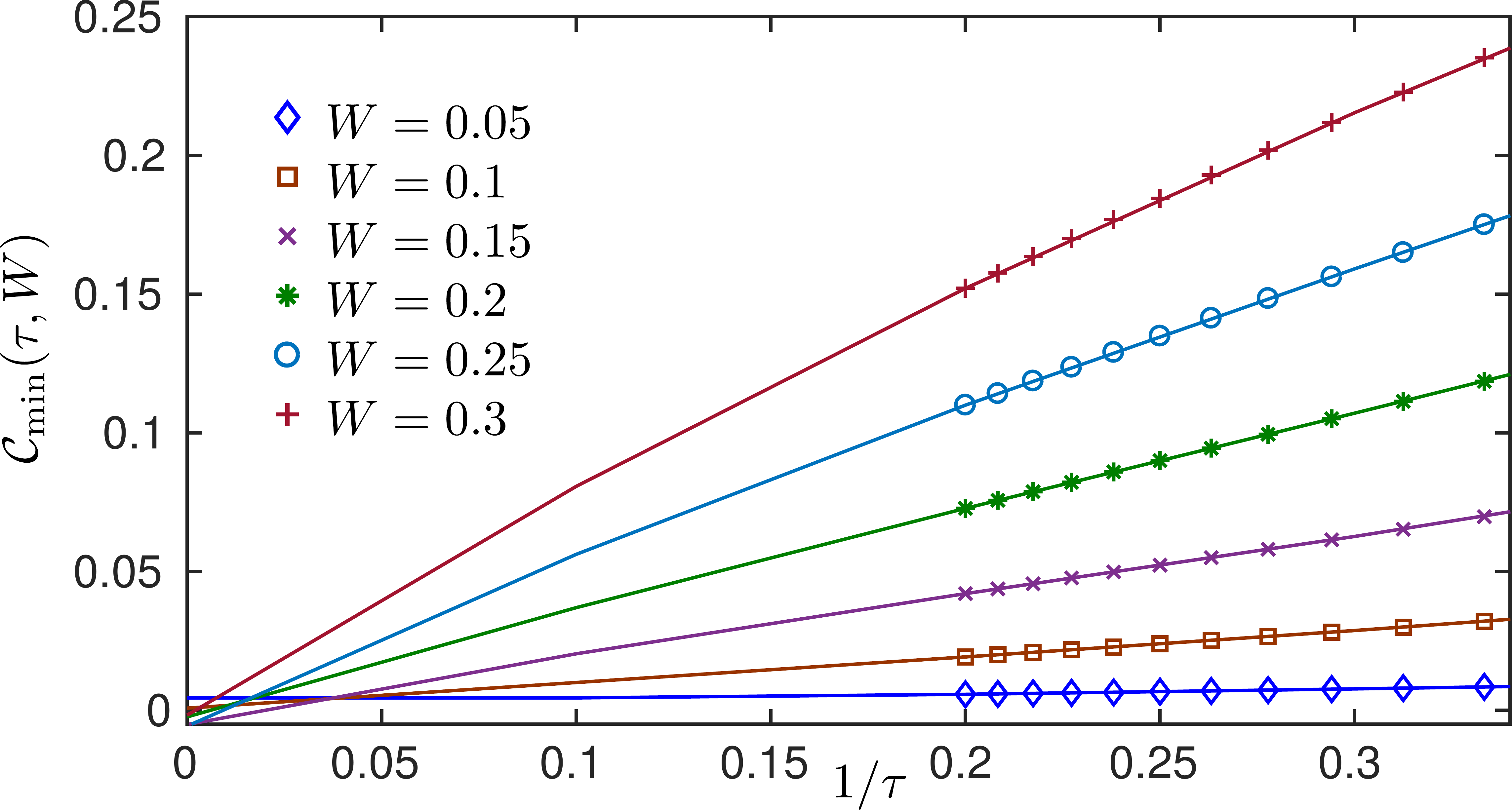}
\caption{The extrapolation of the error function to $\tau\to \infty$, using a fit to a cubic function of $1/\tau$.}
	\label{fig:extrap}
\end{figure}

An important question concerns the fate of the minimum cost function as we increase the total time. In particular, will it saturate to a finite value or decay to zero? We extrapolate the data in the third regime to $\tau\to\infty$ using a polynomial fit to $1/\tau$. The extrapolation (using a cubic polynomial of $1/\tau$) suggests that ${\cal C}_{\rm min}$ likely decays to zero in the limit of infinite time, as shown in Fig.~\ref{fig:extrap}. Upon extrapolation, we find very small \textit{negative} or positive values at $1/\tau=0$, with 0 always contained within the error bars. Note that $\cal C$ is by construction nonnegative. Extrapolation with a fit to a quadratic function of $1/\tau$ yields very small positive values (compared with the error bar). The linear fit is not as high quality, and a quartic fit produces very large error bars. 

Each data point in Fig.~\ref{fig:SA} has a corresponding time-dependent optimal protocol $\Delta_j(t)$. We will present our results on pulse shapes after discussing the properties of the optimal protocols, using the Pontryagin's minimum principle.

\subsection{Optimization with the bang-bang variational ansatz} 
\label{sec:time}

The Monte Carlo simulations show excellent convergence except at the boundary between the first and the second regimes. We believe this is due to the presence of an infinite number of protocols in the first regime and proximity to this degenerate space in the beginning of the second regime. We can improve the results in this region using the bang-bang ansatz. We found that the protocols in the beginning of the second regime have a bang-bang character. This allows us to to use an efficient optimization with the bang-bang ansatz~[\onlinecite{farhi2014,farhi2016,yang2017}] to find the minimum cost function in this regime.

\begin{figure}[]
\includegraphics[width=16cm]{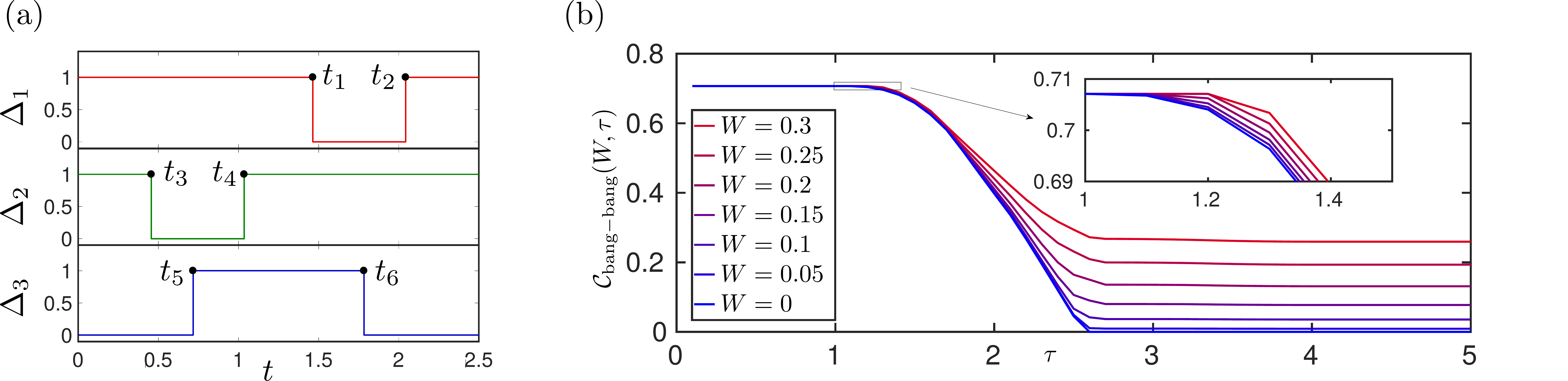}
\caption{(a) The bang-bang ansatz contains six variational parameters $t_1,\dots t_6$. The figure shows an optimal bang-bang protocol for $W=0$ and $\tau=2.5$. (b) The minimum cost function with the six-parameter bang-bang ansatz.  }
	\label{fig:bang}
\end{figure}

The bang-bang ansatz, motivated by Monte Carlo simulations, is shown in Fig.~\ref{fig:bang}(a) and contains six variational parameters. It is easy to write the cost function $\cal C$ as a function of these $t_j$ parameters. We minimized ${\cal C}(t_1,\dots t_6)$ using the interior-point minimization algorithm and found the minimum values of Fig.~\ref{fig:bang}(b). We have better convergence and a lower minimum cost function (than Monte Carlo) at the interface of the first and the second regimes, and comparable performance in the first and the second regime. In the third regime, however, the bang-bang algorithm significantly underperforms the direct simulated-annealing, as the actual optimal protocols become very different from these bang-bang protocols.

\section{Properties of the optimal protocols from Pontryagin's minimum principle}
\label{sec:opt}

We begin by reviewing the Pontryagin's theorem for a generic dynamical system. For dynamical variables $\mathbf{x}(t)=(x_1,x_2,\cdots,x_M)$ driven by control parameters $\mathbf{\Delta}(t)=(\Delta_1,\Delta_2,\cdots,\Delta_N)$ such that
\begin{equation}\label{eq:xmotion}
\partial_t{x}_j={f}_j\big(\mathbf{x},\mathbf{\Delta}\big), \quad \mathbf{x}(t=0)=\mathbf{x}_0,
\end{equation}
we can define conjugate momenta $\mathbf{p}(t)=(p_1,p_2,\cdots,p_M)$, evolving with 
\begin{equation}\label{eq:pmotion}
\partial_t{p}_j=-\mathbf{p}.[\partial_{{x}_j}\mathbf{f}(\mathbf{x},\mathbf{\Delta})],
\end{equation}
where $\mathbf{f}=(f_1,f_2,\cdots, f_M)$. The optimal controls $\mathbf{\Delta}^*$ that minimize a cost function $C[\mathbf{x}(\tau)]$ at the final time $t=\tau$ then satisfy
\begin{equation}\label{eq:pontminexpress}
{\cal H}(\mathbf{p^*,x^*,\Delta^*})=\min_{\mathbf{\Delta}}{\cal H}(\mathbf{p^*,x^*,\Delta})
\end{equation}
for all $0<t<\tau$, where the optimal-control Hamiltonian is defined as
\begin{equation}
{\cal H}(\mathbf{p,x,\Delta})\equiv \mathbf{p}.\mathbf{f(x,\Delta)}.
\end{equation}
The boundary conditions for the conjugate momenta are given by  
\begin{equation}\label{eq:pfinal}
{p_j(\tau)}=\partial_{x_j}C[\mathbf{x}(\tau)].
\end{equation}

In the present problem, we treat the real and imaginary parts of the elements of the density matrix as our dynamical variables. As the density matrix is Hermitian and has trace unity, these dynamical variables are not independent. The independence of the dynamical variables is not a requirement for the theory, however. Let us define
\begin{equation}
\rho_{R}^{\alpha \beta}\equiv  {\rm Re}(\rho^{\alpha \beta}),\quad \rho_{I}^{\alpha \beta}\equiv  {\rm Im}(\rho^{\alpha \beta}).
\end{equation}
We then represent the conjugate momentum to $\rho_{R}^{\alpha \beta}$ ($\rho_{I}^{\alpha \beta}$) by $\Pi_{R}^{\alpha \beta}$ ($\Pi_{I}^{\alpha \beta}$). As shown in Appendix \ref{app:1}, the conjugate momenta satisfy a similar  master equation to that of the density matrix, except the Lindblad double commutator appears with the opposite sign:
\begin{equation}\label{eq:master_conj} 
\partial_t\Pi=\sum_j\left\{i\Delta_j\left[\Pi,O_j\right] + \frac{W^2}{2}\Delta_{j}^2\left[\left [\Pi,O_j\right],O_j\right]\right\}.
\end{equation}
We have combined the conjugate momenta for the real and imaginary parts of $\rho^{\alpha\beta}$ into one complex momentum $\Pi^{\alpha \beta}\equiv \Pi_{R}^{\alpha \beta}+i\Pi_{I}^{\alpha \beta}$. These terms can be collected in a matrix 
\begin{equation}
\Pi=\Pi_R+i\Pi_I.
\end{equation}

The optimal-control Hamiltonian can then be written as
\begin{equation}\label{eq:opt_form} 
{\cal H}=\sum_j\left[F_j(\Pi, \rho)\Delta_j+W^2G_j(\Pi, \rho)\Delta_j^2\right],
\end{equation}
where $F_j$ and $G_j$ are known functions defined in Appendix.~\ref{app:1}.

In the absence of noise, i.e., $W=0$, $\cal H$ is a linear function of $\Delta_j$. Minimizing the optimal-control Hamiltonian \eqref{eq:opt_form} at a given time [see Eq.~\eqref{eq:pontminexpress}] gives
\begin{equation}\label{eq:bang}
\Delta_j(t)=\left\{\begin{array}{c}
0,\quad F_j[\Pi(t),\rho(t)]>0, \\ 
1,\quad F_j[\Pi(t),\rho(t)]<0.
\end{array} \right.
\end{equation}
If $F_j$ identically vanishes over a finite interval, we have a \textit{singular} segment and the optimal protocol can take intermediate values. Otherwise, we have a bang-bang protocol, with the control parameter equal to either its minimum or its maximum allowed value. For $W>0$, there is a third possibility for the minimum, which can occur at 
\begin{equation}\label{eq:smooth}
\Delta_j^{\rm d}\equiv-{F_j\over 2W^2G_j}.
\end{equation}
If $0\leqslant \Delta_j^{\rm d}\leqslant 1$, we need to see which possible value of $\Delta_j$, $(0,1,\Delta_j^{\rm d})$ minimizes $\cal H$, i.e., correspond to $\min(0,F_j+W^2G_j,-F_j^2/4W^2G_j)$. If $\Delta_j^{\rm d}$ is outside this range, we just need to compare the boundary values at $\Delta_j=0,1$.
For non-zero values of noise, the nonlinearity above can lead to protocols that are not bang-bang, even in the absence of singular segments.

The protocols $\Delta_j(t)$ uniquely determines both $\rho$ and $\Pi$ (and consequently $F_j$ and $G_j$) as a function of time. The initial condition for $\rho$ is known and the deterministic master equation yields $\rho(t)$ for all later times. In particular, for a piecewise-constant protocol, using Eq.~\eqref{eq:master3}, we can write
\begin{equation}
\rho(M\Delta t)=e^{K_M(W^2)\Delta t}\dots e^{K_1(W^2)\Delta t}\rho_0.
\end{equation}

For the conjugate momenta $\Pi$, we similarly know the equations of motion \eqref{eq:master_conj}. However, we still need the boundary conditions to determine $\Pi$ for all times. We can get these boundary conditions from Eq.~\eqref{eq:pfinal}, which, in the present problem, gives
\begin{equation}\label{eq:bc_exp}
\Pi^{\alpha\beta}_{R,I}(\tau)={\partial {\cal C}\over \partial \rho^{\alpha \beta}_{R,I}}\Bigg|_{\rho(\tau)}.
\end{equation}
Explicit expressions for the above gradients are given in Eqs.~\eqref{eq:bc_r} and \eqref{eq:bc_i} of Appendix \ref{app:2}. As discussed in Appendix \ref{app:2}, it is important to slightly generalize the cost function so that the above gradients are real. Again, solving the deterministic master equation~\eqref{eq:master_conj} backward in time yields $\Pi(t)$. For the piecewise-constant case, we have
 \begin{equation}
\Pi(M\Delta t)=e^{-K_{M+1}(-W^2)\Delta t}\dots e^{-K_{N-1}(-W^2)\Delta t}e^{-K_N(-W^2)\Delta t}\Pi(\tau),
\end{equation}
where the negative sign for $W^2$ in $K(-W^2)$ follows from Eq.~\eqref{eq:master_conj}.

%
%
%

\section{verification and improvement of optimal protocols using pontryagin's minimum principle}
\label{sec:analyticprotocols}

\subsection{First regime} 
\label{sec:first}
\begin{figure}[t]
\includegraphics[width=17cm]{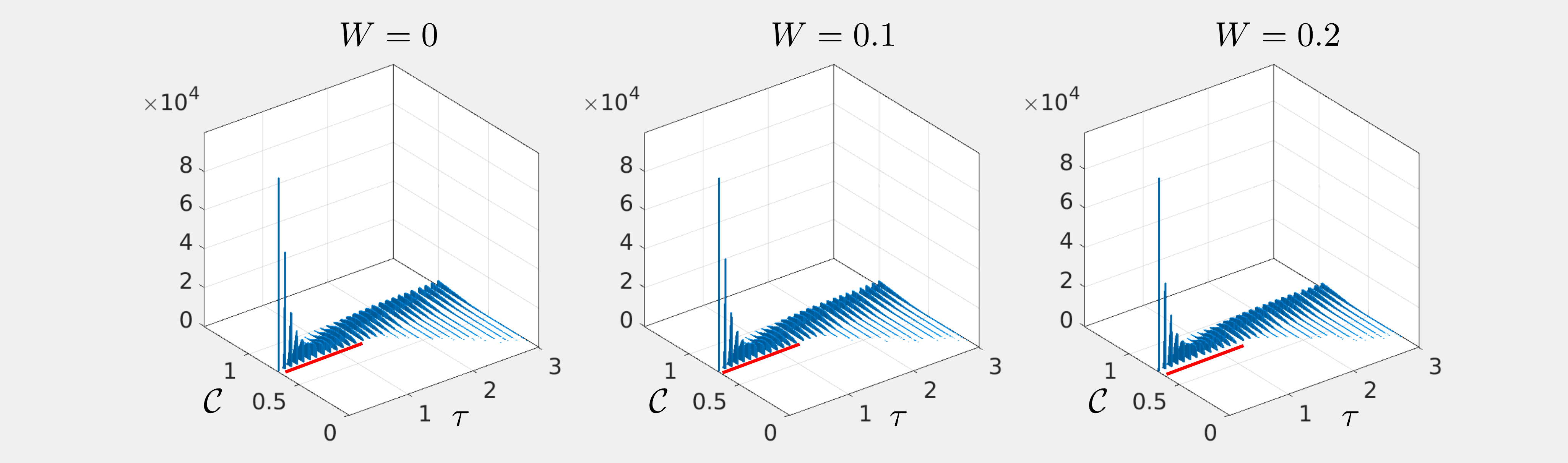}
\caption{Numerically generated approximate histogram of $\cal C$ for randomly generated protocols. Below a critical total time, no random protocol seems to decrease $\cal C$, as indicated by the red lines at the initial $\cal C$.}
	\label{fig:hist}
\end{figure}
In the first regime, the simulations converged to many different protocols, all of which have the same cost function as the identically vanishing protocol
\begin{equation}
\Delta_1(t)=\Delta_2(t)=\Delta_3(t)=0,
\end{equation}
independent of noise.
To check this result, we created a large number random permissible protocols, without optimizing the variational parameters. We then plotted the resulting cost values in a two-dimensional histogram, shown in Fig.~\ref{fig:hist}. We found that for all times in the first regime, the lower limit to the cost function coincided with the cost for $\tau=0$. While the reachable set grows with increasing $\tau$, the growth is away from the target density matrix.

\subsection{Second regime} 
\label{sec:second}
\begin{figure}[]
\includegraphics[width=12cm]{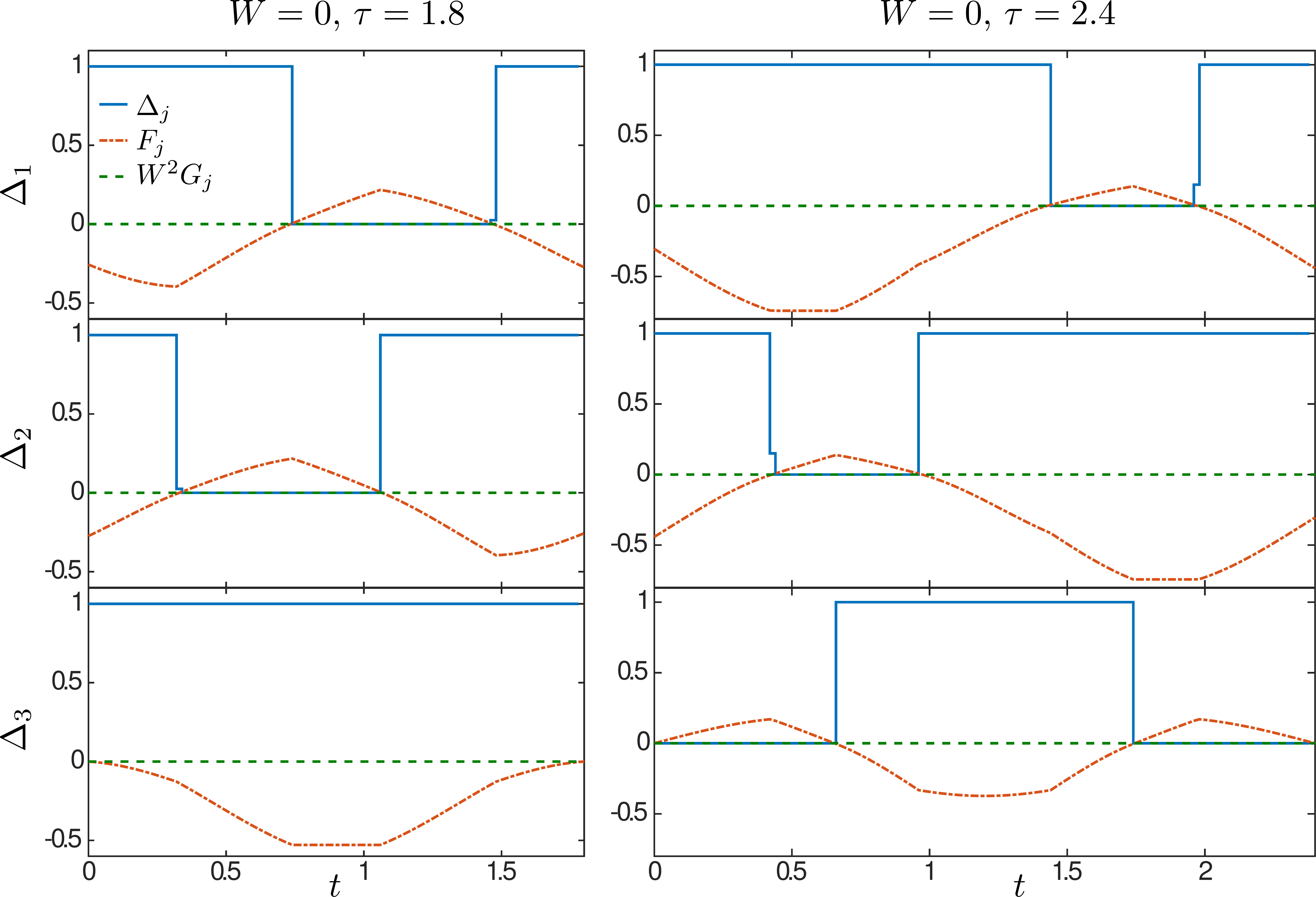}
\caption{Optimal protocols for $W=0$ (solid blue lines), obtained from Monte Carlo simulations, have a bang-bang character with a maximum of two switchings per protocol. $\Delta_j$ is determined by ${\rm sgn}(F_j)$ according to  Eq.\eqref{eq:bang}.}
	\label{fig:pontry1}
\end{figure}
\begin{figure}[]
\includegraphics[width=12cm]{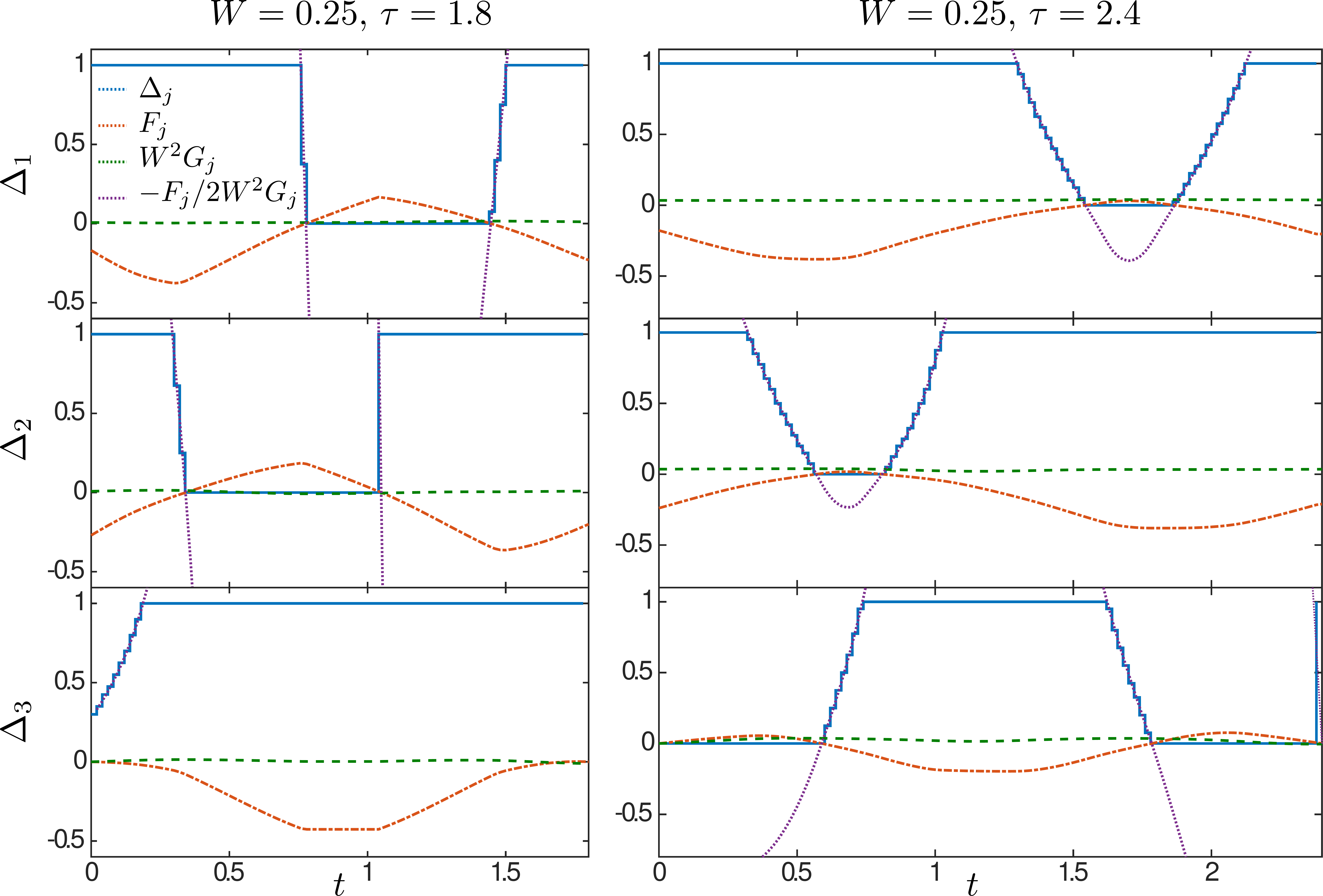}
\caption{Optimal protocols for $W=0.25$ (solid blue lines), obtained from Monte Carlo simulations, have constant pieces with $\Delta_j$ determined by ${\rm sgn}(F_j)$ according to  Eq.\eqref{eq:bang}, as well as continuous pieces in agreement with Eq.~\eqref{eq:smooth}. There is remarkable qualitative similarity with the bang-bang protocols with $W=0$ but the sudden jumps are replaced by smooth segments.}
	\label{fig:pontry2}
\end{figure}

Let us first focus on the noiseless case $W=0$. In this case, we expect bang-bang protocols from the application of Pontryagin's minimum principle [see Eq.\eqref{eq:bang}]. Indeed, we find this behavior in out Monte Carlo results. In Fig.~\ref{fig:pontry1}, we show two representative protocols in the second regime for $W=0$. The optimal $\Delta_j$ is determined by ${\rm sgn}(F_j)$. For shorter times we see that $\Delta_3(t)=1$ while $\Delta_2$ and $\Delta_1$ have two switchings at intermediate times. For longer times, all $\Delta_j$ control parameters have two switchings. The times of these jumps are in agreement with Eq.\eqref{eq:bang}.

Upon turning on the noise, we find similar protocols with segments where $\Delta_j$ is either 0 or 1. However, the sudden jumps get replaced by continuous segments, which become smoother upon increasing either $\tau$ or $W$. These continuous pieces are in full agreement with Pontryagin's minimum principle. They minimize $\cal H$ and are given by Eq.~\eqref{eq:smooth}. 

%

This remarkable agreement indicates that the Monte Carlo simulations have indeed converged to the optimal protocol despite their approximate discrete nature. The almost overlapping continuous pieces from Eq.~\eqref{eq:smooth} are closer to the true optimal protocol than the piecewise-constant Mote Carlo result.  As a given protocol determines $F_j$ and $G_j$, and, consequently, a new protocol containing 0, 1 or $-F_j/2W^G_j$, this improvement can be iterated to arbitrary precision, leading to numerically exact optimal protocols. We note that although our Monte Carlo simulations are very accurate, the Pontryagin theorem is a powerful practical tool in finding the optimal protocol from a much less accurate simulation (or other minimization routine), which produces an approximate optimal protocol. If the approximate protocol is not too far from optimal, we have found that the above-mentioned iteration of Pontryagin's theorem can yield the optimal protocol.

\subsection{Third regime} 
\label{sec:third}

\begin{figure}[]
\includegraphics[width=16cm]{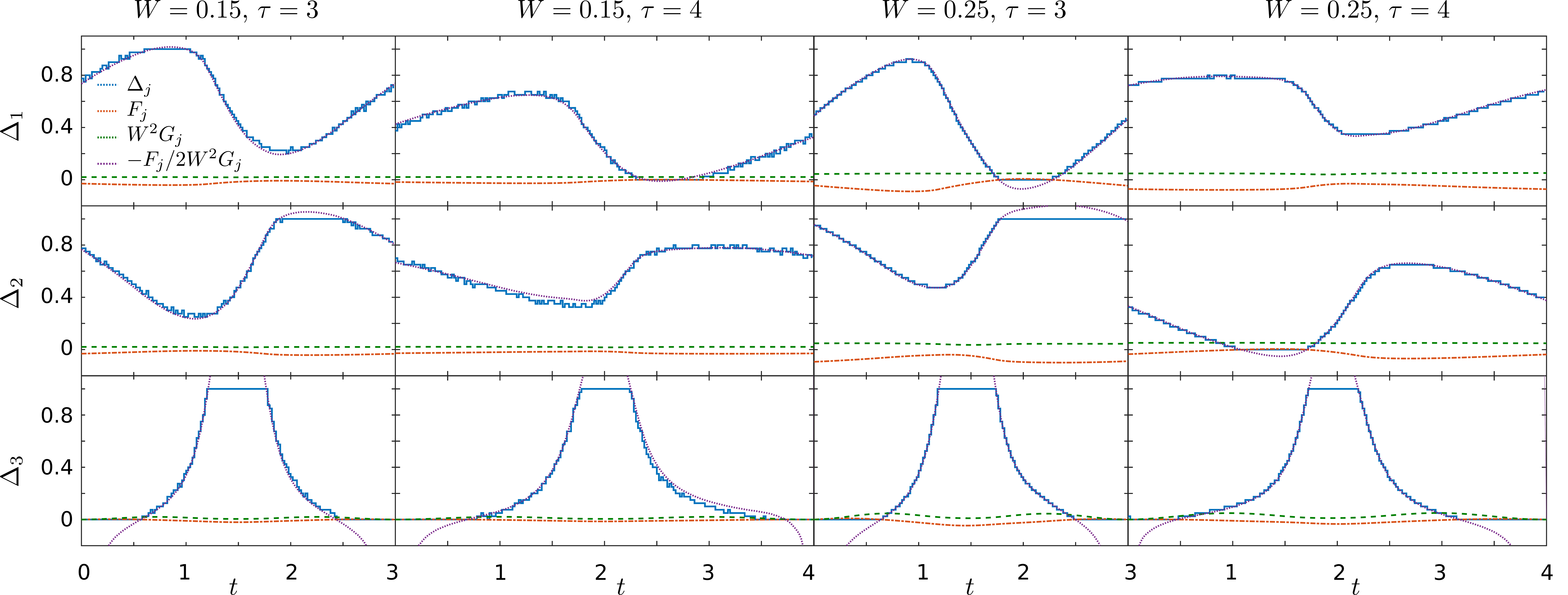}
\caption{The general pattern of the optimal protocols in the third regime. All protocols are in perfect agreement with Pontryagin's minimum principle.}
	\label{fig:pontry3}
\end{figure}

In the third regime, the protocols become more and more continuous. As shown in Fig.~\ref{fig:pontry3}, the general pattern is as follows. For $\Delta_1$, we have an S-shaped protocol (ascending, descending and ascending again). Upon increasing the total time, the pulse shape is squished in the vertical direction. When this S-shaped pulse is negative, the optimal protocol has a constant segment with $\Delta_1=0$. For $\Delta_2$, there is a similar S-shaped protocol (now descending, ascending, and descending again). When this S-shaped pulse is negative (larger than 1), the optimal protocol has a constant segment with $\Delta_2=0(1)$. $\Delta_3$ has a fixed pattern shown in Fig.~\ref{fig:pontry3}.

\section{conclusions}
\label{sec:conclusion}
Motivated by an antiadiabatic behavior in the presence of high-frequency noise, which may be detrimental for all adiabatic quantum information processing schemes, we investigated optimal shortcuts to adiabatic evolution in the presence of white noise for a particular Majorana-based gate.
We formulated the noise-averaged dynamics in terms of a master equation and performed Monte Carlo simulations to minimize a cost function of the final density matrix, which vanishes for perfectly adiabatic evolution in the absence of noise. 

We found three distinct regimes as a function of the total time. As we are interested in small cost functions and precise gates, the third regime is of particular interest. We found that this optimization completely eliminates the antiadiabatic behavior.

We discovered very rich behavior in the pulse shapes of our optimal protocols. Starting with bang-bang protocols for vanishing noise, we found that the optimal protocols developed continuous segments as noise was increased. We performed an in-depth analysis of the application of the Pontryagin's minimum principle, which not only provides a mathematical understanding of the pulse shapes, but also allows us to arbitrarily improve approximate numerically obtained optimal protocols. This method may be of value to various quantum optimization schemes and variational quantum algorithms. 

With the rapid development of quantum technologies, striking a balance between performance and robustness is crucially important. Topological quantum computing relies on adiabaticity and offers remarkable robustness. However, topological protection can be lost in the limit of long times due to the presence of noise or other factors like quasiparticle poisoning. Finding fast dynamical protocols that cancel the effects of noise suggest that even without strict topological protection, we may be able to perform  high-performance and reliable information processing with Majorana-based platforms.

\acknowledgements{A.R. is grateful to Marcel Franz and Babak Seradjeh for collaboration in a related project. This work was supported by WWU College of Science and Engineering.}
\appendix
\section{Optimal-control Hamiltonian and the master equations for the conjugate momenta}\label{app:1}
We start by expanding the master equation in terms of the components as
\begin{equation}
\partial_t\rho^{\alpha\beta}=\sum_j\left[i\Delta_j\left(\rho^{\alpha\gamma}O_j^{\gamma \beta}-O_j^{\alpha\gamma}\rho^{\gamma\beta}\right)-W^2\Delta_j^2\rho^{\alpha\beta}+W^2\Delta_j^2\left(O_j^{\alpha\gamma}\rho^{\gamma\sigma}O_j^{\sigma\beta}\right)\right],
\end{equation}
where summation over repeated indices is implied. We have used the identity $O_j^2=\openone$, to write $\left[\left [\rho,O_j\right],O_j\right]=2\left(\rho-O_j\rho O_j\right)$. Noting that $O_1$ is imaginary and $O_2$ and $O_3$ are real, we can break the above equation into real and imaginary parts and construct the optimal-control Hamiltonian as
\begin{equation}\label{eq:opt_H}
\begin{split}
{\cal H}&=\Pi_R^{\alpha\beta}\left[i\Delta_1\left(\rho^{\alpha\gamma}_RO_1^{\gamma \beta}-O_1^{\alpha\gamma}\rho^{\gamma\beta}_R\right)
-\sum_{j=2}^3\Delta_j\left(\rho^{\alpha\gamma}_IO_j^{\gamma \beta}-O_j^{\alpha\gamma}\rho^{\gamma\beta}_I\right)
-W^2\left(\sum_{j=1}^3\Delta_j^2\right)\rho^{\alpha\beta}_R
+W^2\left(\sum_{j=1}^3\Delta_j^2O_j^{\alpha\gamma}O_j^{\sigma\beta}\right)\rho^{\gamma\sigma}_R\right]\\
&+\Pi_I^{\alpha\beta}\left[i\Delta_1\left(\rho^{\alpha\gamma}_IO_1^{\gamma \beta}-O_1^{\alpha\gamma}\rho^{\gamma\beta}_I\right)
+\sum_{j=2}^3\Delta_j\left(\rho^{\alpha\gamma}_RO_j^{\gamma \beta}-O_j^{\alpha\gamma}\rho^{\gamma\beta}_R\right)
-W^2\left(\sum_{j=1}^3\Delta_j^2\right)\rho^{\alpha\beta}_I
+W^2\left(\sum_{j=1}^3\Delta_j^2O_j^{\alpha\gamma}O_j^{\sigma\beta}\right)\rho^{\gamma\sigma}_I\right].
\end{split}
\end{equation}

The equations of motion for the conjugate momenta can be written as 
\begin{eqnarray}
\partial_t\Pi_R^{\lambda\mu}=-{\partial{\cal H}\over \partial \rho_R^{\lambda \mu}}, \qquad \partial_t\Pi_I^{\lambda\mu}=-{\partial{\cal H}\over \partial \rho_I^{\lambda \mu}},
\end{eqnarray}
which leads to
\begin{equation}
\partial_t\Pi^{\lambda\mu}=-i\Delta_1\left(\Pi^{\lambda\beta}O_1^{\mu\beta}-O_1^{\beta\lambda}\Pi^{\beta\mu}\right)+i\sum_{j=2}^3\Delta_j\left(\Pi^{\lambda\beta}O_j^{\mu\beta}-O_j^{\beta\lambda}\Pi^{\beta\mu}\right)+W^2\left(\sum_{j=1}^3\right)\Pi^{\lambda\mu}-W^2\left(\sum_{j=1}^3\Delta_j^2O_j^{\alpha\lambda}O_j^{\mu\beta}\right)\Pi^{\alpha\beta}.
\end{equation}
Noting that all $O_j$ matrices are Hermitian, $O_1$ is imaginary and $O_{2,3}$ are real, we have
\begin{equation}
O_1^{\alpha\beta}=-O_1^{\alpha\beta},\qquad
O_2^{\alpha\beta}=O_2^{\alpha\beta},\qquad
O_3^{\alpha\beta}=O_3^{\alpha\beta}.
\end{equation}
The above relationships immediately lead to Eq.~\eqref{eq:master_conj}.
We can also read off $F_j$ and $G_j$ in Eq. \eqref{eq:opt_form} from Eq.~\eqref{eq:opt_H} as follows:
\begin{eqnarray}
F_1&=& i\Pi_R^{\alpha\beta}\left(\rho^{\alpha\gamma}_RO_1^{\gamma \beta}-O_1^{\alpha\gamma}\rho^{\gamma\beta}_R\right)+i\Pi_I^{\alpha\beta}\left(\rho^{\alpha\gamma}_IO_1^{\gamma \beta}-O_1^{\alpha\gamma}\rho^{\gamma\beta}_I\right),\\
F_j&=&-\Pi_R^{\alpha\beta}\left(\rho^{\alpha\gamma}_IO_j^{\gamma \beta}-O_j^{\alpha\gamma}\rho^{\gamma\beta}_I\right)+\Pi_I^{\alpha\beta}\left(\rho^{\alpha\gamma}_RO_j^{\gamma \beta}-O_j^{\alpha\gamma}\rho^{\gamma\beta}_R\right),\quad j=2,3,\\
G_j&=&-\Pi_R^{\alpha\beta}\rho_R^{\alpha\beta}-\Pi_I^{\alpha\beta}\rho_I^{\alpha\beta}+O_j^{\alpha\gamma}O_j^{\sigma\beta}\left(\Pi_R^{\alpha\beta}\rho^{\gamma\sigma}_R+\Pi_I^{\alpha\beta}\rho^{\gamma\sigma}_I\right), \quad j=1,2,3.
\end{eqnarray}

\section{Gradient of the cost function and the boundary cndition for the conjugate momenta}\label{app:2}
As in Eq.~\eqref{eq:bc_exp}, the boundary conditions for the conjugate momenta rely on the derivatives of the cost function with respect to the real and imaginary parts of the elements of the density matrix. The cost function, written as $C[\sigma,\rho(\tau)]\equiv \frac{1}{2}{\rm tr}\sqrt{\left[\sigma-{\rho}(\tau)\right]^2}$ in Eq.~\eqref{eq:cost}, makes use of the assumption that the density matrices $\sigma $ and $\rho$ are Hermitian. In our formulation, we treated the real and imaginary parts of the individual elements of the density matrix $\rho$ as dynamical variables. These elements are not independent due to the fact that the density matrix is Hermitian and has trace one. However, if we compute the variation of $C(\sigma,\rho)$ with respect to one element of the density matrix, we are not guaranteed to obtain real values, as the variation violates the Hermiticity constraint. We can reduce the number of dynamical variables and only use variations of the density matrix that are Hermitian. Alternatively we can make use of a more general expression for the cost function
\begin{equation}
C[\sigma,\rho(\tau)]\equiv \frac{1}{2}{\rm tr}\sqrt{\left[\sigma-{\rho}(\tau)\right]^\dagger\left[\sigma-{\rho}(\tau)\right]},
\end{equation}
which yields real values even for non-Hermitian $\rho$ and the same values as Eq.~\eqref{eq:cost} for a Hermitian $\rho$.  We take the latter approach here.

Let us begin by computing the derivative of the cost function with respect to the real part, $\rho^{ij}_R$ , of an element of the density matrix. We add an infinitesimal matrix $\delta \rho$ with
\begin{equation}\label{eq:drho}
(\delta \rho)_{\alpha \beta}=\delta_{\alpha i}\delta_{\beta j}\epsilon
\end{equation}
to the density matrix.
The variation of the cost function is obtained from ${\cal C}+\delta {\cal C}= \frac{1}{2}{\rm tr}\sqrt{\left(\sigma-{\rho}-\delta\rho\right)^\dagger\left(\sigma-{\rho}-\delta\rho\right)}\approx  \frac{1}{2}{\rm tr}\sqrt{(\sigma-\rho)^2-(\delta\rho)^\dagger(\sigma-\rho)-(\sigma-\rho)\delta\rho}$,
where we have made use of the Hermiticity of $\sigma$ and $\rho$, but allowed $\delta \rho$ to be non-Hermitian. If we now define a matrix $A$ with
\begin{equation}\label{eq:A_mat}
A_{\alpha \beta}\equiv\delta_{\alpha j}(\sigma-\rho)_{i \beta}+(\sigma-\rho)_{\alpha i }\delta_{\beta j},
\end{equation}
we have ${\cal C}+\delta {\cal C}\approx  \frac{1}{2}{\rm tr}\sqrt{(\sigma-\rho)^2-A\epsilon}$.
The matrix $A$ is constructed by taking a matrix of zeros, and respectively adding the $i$th row and column of the matrix $(\sigma-\rho)$ to its $j$th row and column. 

We can further define a matrix $X$ through $\sqrt{(\sigma-\rho)^2-A\epsilon}=\sqrt{(\sigma-\rho)^2}+\epsilon X$.
Squaring both sides gives $X\sqrt{(\sigma-\rho)^2}+\sqrt{(\sigma-\rho)^2}X=-A$, which for an invertible $\sqrt{(\sigma-\rho)^2}$ leads to 
\begin{equation}
X+\sqrt{(\sigma-\rho)^2}X\left[\sqrt{(\sigma-\rho)^2}\right]^{-1}=-A\left[\sqrt{(\sigma-\rho)^2}\right]^{-1}.
\end{equation}
Using the cyclic property of the trace, we find from ${\delta {\cal C} }={1\over 2}\epsilon{\rm tr}(X)$ that
\begin{equation}\label{eq:bc_r}
{\partial {\cal C}\over \partial [ \rho_R^{ij}]}=-{1\over 4}{\rm tr}\left\{A\left[\sqrt{(\sigma-\rho)^2}\right]^{-1}\right\},
\end{equation}
where the matrix $A$ depends on $i$ and $j$ through Eq.~\eqref{eq:A_mat}. A similar argument gives
\begin{equation}\label{eq:bc_i}
{\partial {\cal C}\over \partial [ \rho_I^{ij}]}=-{i\over 4}{\rm tr}\left\{B\left[\sqrt{(\sigma-\rho)^2}\right]^{-1}\right\}, \qquad B_{\alpha \beta}\equiv-\delta_{\alpha j}(\sigma-\rho)_{i \beta}+(\sigma-\rho)_{\alpha i }\delta_{\beta j},
\end{equation}
where the matrix $B$ defined above also depends on $i$ and $j$.


\bibliography{maj_optimal_noise.bib}

\end{document}